\documentclass[twocolumn,amsmath,amssymb,english,superscriptaddress]{revtex4-1}
\usepackage{graphicx,natbib,subfigure,tabularx,epsfig,longtable,amsfonts,rotating,subfigure,hyperref,babel,currfile}
\usepackage{color}

\newcommand{\bea}{\begin{eqnarray}}
\newcommand{\eea}{\end{eqnarray}}

\newcommand{\be}{\begin{equation}}
\newcommand{\ee}{\end{equation}}


\begin{document}
\title{Optical Schr\"odinger Cat States in One Mode and
	Two Coupled-Modes Subject to Environments}
\author{X. L. Zhao}
\affiliation{School of Physics and Optoelectronic Technology, Dalian
	University of Technology, Dalian 116024, China\\}
\author{Z. C. Shi}
\affiliation{Department of Physics, Fuzhou University, Fuzhou 350002, China\\}
\author{M. Qin}
\affiliation{School of Physics and Optoelectronic Technology, Dalian
	University of Technology, Dalian 116024, China\\}
\author{X. X. Yi}\thanks{ E-mail: yixx@nenu.edu.cn}
\affiliation{Center for Quantum Sciences and School of Physics, Northeast Normal University, Changchun 130024, China\\}
\date{\today}
\begin{abstract}
Taking the decoherence into account, we investigate nonclassical features of the optical Schr\"odinger cat states  in  one mode and two coupled-modes
systems with two-photon driving. In the one mode system, the relationship between the Schr\"odinger cat states and the system parameters is derived.
We observe that in the presence of single-photon decay the steady states would be a mixture of Schr\"odinger cats. The dynamics and steady states of
such a cat versus single-photon decay are examined.  In the two coupled-modes cases with linear and nonlinear couplings, the dynamics of entanglement
and mutual information are examined with two different initial states and single-photon decay. Compared to the linear coupling case, more complicated
structure appears in the Wigner function in the nonlinear coupling case. The joint quadrature distributions  are also explored. Such nonclassical states can
 be used not only in exploring the boundary  between the classical and the quantum worlds  but also in quantum metrology  and quantum information processing.
\end{abstract}
\maketitle
\section{introduction}\label{introduction}
Schr\"odinger cat state has captured plenty of attention since it was first proposed by Schr\"odinger in the early days of quantum mechanics.
 Such kinds of macroscopic superposition states have been explored in various systems. For example, it has been studied in an ensemble
 of two-level atoms coupled  with a cavity mode\cite{pra562249} and observed  in
 superconducting quantum interference devices\cite{prb577474,nature40643,squidcat}. In quantum Ising model, topological defects can be
 put in a non-local superposition--a topological Schr\"odinger cat state \cite{naturephysics849}. And in qubit-oscillator systems, a scheme
 based on conditional displacement is proposed to create Schr\"odinger cat states by coupling the quantum bit and harmonic oscillator\cite{pra93033853}.
 Not only two-component Schr\"odinger cat states but also multicomponent Schr\"odinger kittens can be generated \cite{pla19885,oc132452,nature495205}.
Besides, by designed double-photon processes, a Schr\"odinger cat state can be confined in a box\cite{science347853}.

In quantum optics, coherent state is a state that is very close  to macroscopic state\cite{scully}. Thus the superposition of two coherent states
 $|\beta_1\rangle$ and $|\beta_2\rangle$ (not overlapping) can be regarded as the  Schr\"odinger cat state, i.e.,  $|\phi_{cat}\rangle=p_1|\beta_1\rangle+p_2|\beta_2\rangle$, $|p_1|^2+|p_2|^2=1$
where $p_1$, $p_2$, $\beta_1$ and $\beta_2$ are all complex. Usually the superpositions of two coherent states out of $180^{\circ}$
in phase are particularly interesting, and the superpositions of more than two coherent states would termed as Schr\"odinger kittens \cite{pra87042315}.
Since the cat states can be composed of coherent states in quantum optics, several works have been devoted to explore the  properties of these
states such as squeezing, photon anti-bunching and non-Poisson distribution~\cite{prl5713,pra415261,pra436458,pra456570,pra47552,pra475024,pra484062,pra49490,pra511698}.
The Schr\"odinger cat states can be produced in double-photon driven-dissipative system~\cite{jmo401053}, where the dissipation has negative
 effect on preparing these states \cite{pra312403}.  Further more, in the presence of single- and double-photon absorption and emission, exact
 stationary  solutions for the diagonal elements of the master equation has been found \cite{aps48349}.

In this work, we explore how the Schr\"odinger cat states emerge  in the double-photon pumping and absorbing (decay) processes
with single-photon decay in one- and two coupled-modes systems. Such cat states can be regarded as a result of   competition between the double-photon
generation and destruction\cite{pla174185}.  Without single-photon decay, the steady state would be a cat state with parity identical to
the initial one. When the initial state is a superposition of Fock states with opposite parity, the steady state would be a weighted mixture of even
and odd Schr\"odinger cats, and the weight of each component is determined by  the initial condition. Such a relation may be revealed by constructing
conserved quantities\cite{njp16045014}.  While intriguing properties emerge in the double-photon driven-dissipative process, single-photon
decay is usually inevitable. Further investigation shows that the decay not only diminishes  the negative interference fringes in the Wigner functions
but also leads to linearly decrease of the photon number in the steady state.  These however do not affect the generation of the cat state with  small  single-photon decay rate, and the double-photon driving can prolong the lifetime of the optical cat state.

By extending the one-mode model to two coupled-modes with linear and nonlinear couplings, the dynamical behaviors of entanglement
and mutual information  for two types of initial states are examined  in the presence of single-photon decay. Single-photon decay not only leads to
 the vanishing of the negative regions in the Wigner functions. The entanglement and mutual information are also suppressed by the
 single-photon decay. Compared to the linear coupling case, entanglement and mutual information in the nonlinear coupling case are robust
against the single-photon decay. And multicomponent cat states appear with sub-Planck phase interference structure in the phase space
\cite{PRA78034101,PRA70053813}. Besides being used in exploring the boundary between the classical and the quantum worlds, the cat states can be used  in  quantum  metrology
\cite{Science342607,Nature412712,Science342568,Science3521087,PRA73023803,PRA78034101,PRA70053813,PRA66023819,PRA66023819}
and quantum information processing \cite{nature448784,pra64052308,science31283,arxiv0509137,Science342607}.

This work is organized as follows: in Sec.\ref{r:onemode}, we introduce a model  with double-photon driven-dissipative process for one mode system.
 Then the dynamics of average photon number, parity, entropy and purity  with different single-photon decays are examined. Interference behaviors
 characterized  by Wigner functions and quadrature distributions are calculated  in the presence of single-photon decay for two different initial states.
 In Sec.\ref{r:twomode}, in the cases of two modes with linear and nonlinear couplings,  behavior of entanglement,  mutual information for
 various single-photon decays with  different initial states are shown. The Wigner function and joint quadrature distributions of the steady states
 are also explored. At last, we summarize in Sec.\ref{r:sum}.

\section{one-mode system}\label{r:onemode}

In this case, we consider the situation that the cavity mode with double-photon driving is coupled to a Markovian environment.
The dynamics  of this optical system $\rho$ can be described by the Lindblad master equation,
$\partial_t \rho= i\,[\rho, \hat{H}_a]+(\mathcal{L}^{(1)}_a+\mathcal{L}^{(2)}_a) \rho$ ($\hbar=1$ hereafter). The operator $\hat{H}_a$
is the system Hamiltonian and $\mathcal{L}^{(1)}_a$ and $\mathcal{L}^{(2)}_a$ denote the single-photon and double-photon Lindblad
dissipation super-operators. The double-photon process can be realized by Kerr effect in Josephson junctions~\cite{science347853}. Then we have
\begin{eqnarray}
\begin{aligned}
&\hat{H}_a=-\Delta_a\hat{a}^\dagger \hat{a}+\frac{U_a}{2}\hat{a}^{\dag2} \hat{a}^2+\frac{G_a}{2} \hat{a}^2 + \frac{G_a^{\ast}}{2} \hat{a}^{\dag2} , \\
& \mathcal{L}_{a}^{(1)} \rho=\frac{\gamma_a}{2}\left(2 \hat{a} \rho \hat{a}^{\dag} -\hat{a}^{\dag}\hat{a} \rho - \rho \hat{a}^{\dag} \hat{a} \right),  \\
&\mathcal{L}_{a}^{(2)} \rho=  \frac{\eta_a}{2}\left(2 \hat{a}^2 \rho \hat{a}^{\dag2} -\hat{a}^{\dag2}\hat{a}^2 \rho - \rho \hat{a}^{\dag2}\hat{a}^2\right),
\end{aligned}\label{Hs}
\end{eqnarray}
where $\Delta_a$ is the pump-cavity detuning ($\Delta_a$=0 following) and $U_a$ is the photon self-interaction strength. $G_a$ is proportional
to the degenerate double-photon parametric susceptibility. $\gamma_a$ denotes the strength for the single-photon dissipation and $\eta_a$ is the
double-photon decay rate related to the cross-section for double-photon absorption and the single-photon decay destroys the coherence in one-mode
 case. In two modes system, the single-photon decay would influence the coupled modes through the coupling interactions. Without single-photon decay,
 the steady states would be semi-deterministic Schr\"odinger cat states related to the initial conditions. The optical Schr\"odinger cat states are generated
 from the competition between the double-photon parametric and destructive processes.

Define the elements of the density matrix $\langle m|\rho|n\rangle=\rho_{m,n}=\sqrt{\frac{m!}{n!}}\phi_{m,n}$,
where $|m\rangle$ and $|n\rangle$ are the Fock states, the density matrix elements $\phi_{m,n}$ satisfy the coupled differential equations,
\begin{eqnarray}
\begin{aligned}
\partial_{t}\phi_{m,n}=&-i[\frac{U_a}{2}(m+n-1)(m-n)\phi_{m,n}\\
&+\frac{G_a}{2}((m+1)(m+2)\phi_{m+2,n}-n(n-1)\phi_{m,n-2})\\
&+\frac{G_a^{\ast}}{2}(\phi_{m-2,n}-\phi_{m,n+2})] \\
&+\frac{\gamma_a}{2}[2(m+1)\phi_{m+1,n+1}-(m+n)\phi_{m,n}]\\
& +\frac{\eta_a}{2}[2(m+1)(m+2)\phi_{m+2,n+2}\\
&-(m(m-1)+n(n-1))\phi_{m,n}].
\end{aligned}\label{Hs}
\end{eqnarray}
The dynamics and steady states can be obtained by solving these equations numerically in a truncated Fock state space.

\subsection{Double-photon driven-dissipative process without single-photon
decay}\label{r:initial}

First, we consider the  system undergoing a double-photon driven-dissipative  process but with non single-photon decay ($\gamma_a=0$). In this
 case, the dynamics and steady-state not only depend on the system parameters but also on the initial state since the double-photon processes link
 two states of the same parity. One can obtain  the steady state by setting
 $\partial_t \rho= i\,[\rho, \hat{H}_a]+(\mathcal{L}^{(1)}_a+\mathcal{L}^{(2)}_a) \rho=0$ with $\gamma_a=0$.
\begin{eqnarray}
\begin{aligned}
\rho_{ss}=& p_{++}|\alpha+\rangle\langle\alpha+|+p_{--}|\alpha-\rangle\langle\alpha-|\\
&+p_{+-}|\alpha+\rangle\langle\alpha-|+p_{-+}|\alpha-\rangle\langle\alpha+|,
\label{rhoss}
\end{aligned}
\end{eqnarray}
where $|\alpha\pm\rangle=\mathcal{N}_{\pm}^{-1}(|\alpha\rangle\pm|-\alpha\rangle)$ are the even and odd Schr\"odinger
 cat states, respectively, and $\mathcal{N}_{\pm}$ are the normalization constant. We obtain the critical parameter `$\alpha$' in (\ref{rhoss}) satisfying,
\begin{eqnarray}
\alpha=\pm\sqrt{\frac{G_a^{\ast}}{ i\eta_a-U_a}},\label{sol}
\end{eqnarray}
where $|\alpha\rangle$ and $|-\alpha\rangle$ are components for the Schr\"odinger cat states. Thus the desired Schr\"odinger cat states may be obtained by adjusting the parameters of the system. Usually large $\alpha$ is desirable since the  nonclassical signatures would be more obvious for Schr\"odinger cat states with large average photon number.

We will use the Wigner function to characterize  nonclassical states in the phase space\cite{pr40749}, which can be reconstructed
by homodyne tomography technique\cite{pra402847}. The appearance of negative value of Wigner function is the necessary condition to indicate the appearance of nonclassical
state~\cite{scully}. It is defined as $W(\alpha)=\frac{1}{\pi}Tr[\rho \hat{D}_\alpha\hat{P}\hat{D}^{\dag}_\alpha]$,  where $\hat{D}_\alpha=e^{\alpha \hat{a}^{\dag}-\alpha^{\ast} \hat{a}}$
is the displacement operator and $\hat{P}=e^{i\pi \hat{a}^{\dag}\hat{a}}$ is the photon parity operator. Different from the P-function,  Wigner function is a quantum analog of the classical
Liouville phase space probability function\cite{scully}. Even though the non-classicality of a states can be witnessed  by P-function,
its delicate mathematical structure makes it hard to be reconstructed by experimental data~\cite{scully}.  For the case of even and odd Schr\"odinger
cat states $|\xi\pm\rangle$, the Wigner function is
\begin{eqnarray}
\begin{aligned}
\vspace{-5em}
W(\alpha)=&\frac{2}{\pi\mathcal{N}_{e(o)}}\{e^{-2|\alpha-\xi|^2}+e^{-2|\alpha+\xi|^2}\\
&+2e^{-2|\alpha|^2}(-1)^{e(o)}\cos[4Im(\alpha^*\xi)]\}.\\
\label{Wignerf}
\vspace{-5em}
\end{aligned}
\end{eqnarray}

\begin{figure}
	\includegraphics[width=0.5\textwidth]{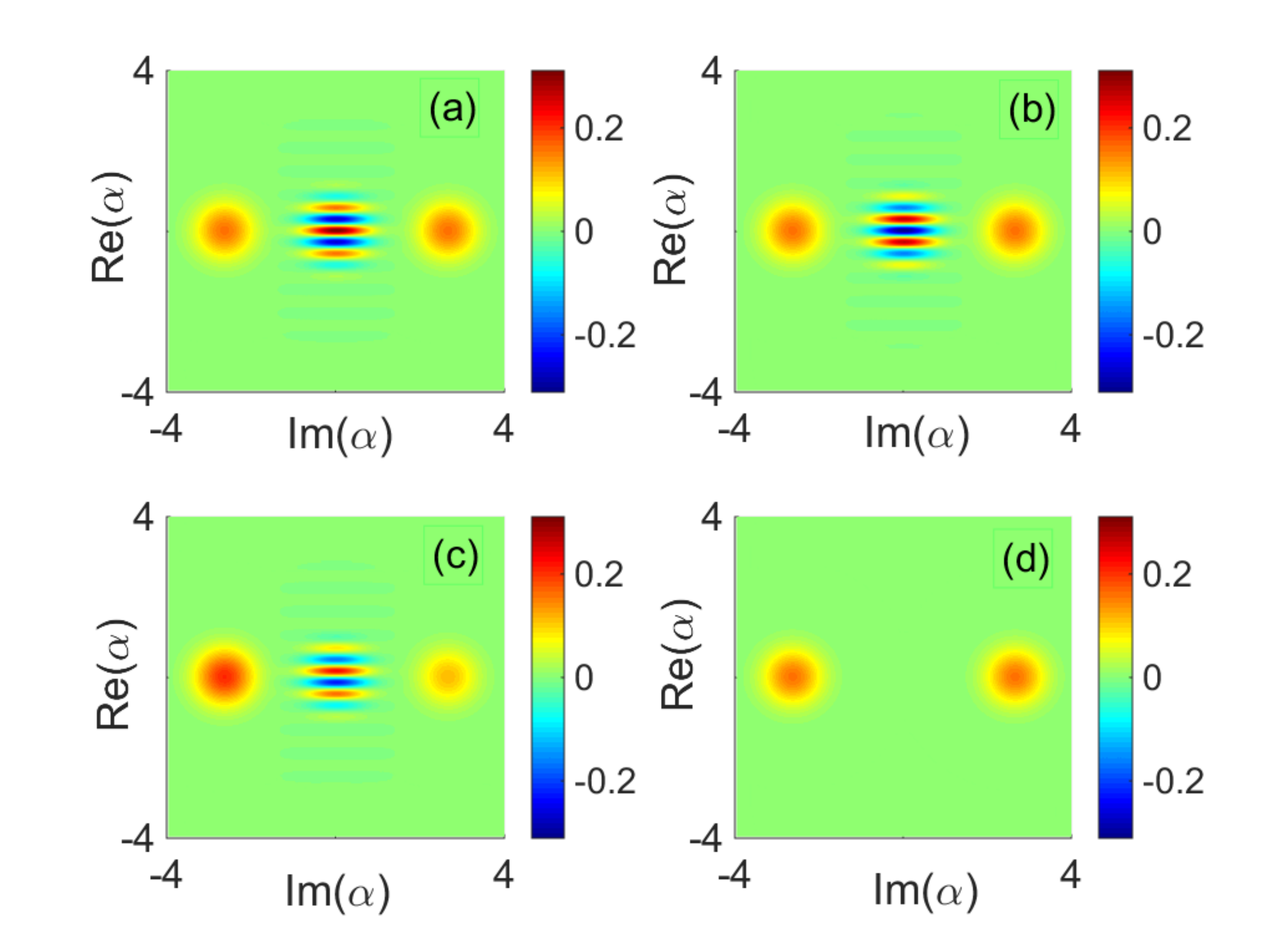}
	\caption{(a)-(d) show the Wigner functions for the steady states with the initial states, $ |0\rangle\langle0|$,  $|1\rangle\langle1|$,
		 $1/2(|0\rangle+|1\rangle)(\langle0|+\langle1|$) and $1/2(|0\rangle \langle0|+|1\rangle\langle1|)$ respectively. (a)Even Schr\"odinger cat state;
		 (b) Odd Schr\"odinger cat state; (c) Superposition of even and odd Schr\"odinger cat states with different ratios; (d) Statistical mixture of even and odd
		 Schr\"odinger cat states; The other parameters are $\Delta_a$=0,  $U_a/\eta$=1, $G_a/\eta_a$=10$e^{-i\pi/4}$ and $\gamma_a$=0.}
	\label{f:initialstate}
\end{figure}
Here $\mathcal{N}_{e(o)}=(1+(-1)^{e(o)} e^{-2|\xi|^2})$ are the constants for even and odd Schr\"odinger cat states and the subscripts $e$ and $o$
denote  even and odd Schr\"odinger cat states. From another point of view, the coherence of the Schr\"odinger cat  state can also be identified by the interference
fringes in quadrature distributions which can be identified by homodyne detection~\cite{prl5713,pra415261}. Such a distribution for a state density $\rho$
can be written as
\begin{eqnarray}
\begin{aligned}
\vspace{-5em}
\mathcal{P}(X)=\langle X,\phi|\rho|X,\phi\rangle,
\label{qp}
\vspace{-5em}
\end{aligned}
\end{eqnarray}
where $|X,\phi\rangle$ is an eigenstate of the quadrature operator $\frac{1}{\sqrt{2}}(\hat{a}^{\dag}e^{-i\phi}+\hat{a}e^{\phi})$.
Based on the relation $\langle X,\phi|n\rangle$= $(2^nn!\sqrt{n})^{-1/2}H_n(X)e^{-X^2/2}e^{-i n\phi}$, we gain the quadrature distribution
of a state, here $H_n(X)$ are the Hermite polynomials of order $n$. Interference patterns in the quadrature distributions is also a signature of nonclassical
properties.
\begin{figure}
	\includegraphics[width=0.48\textwidth]{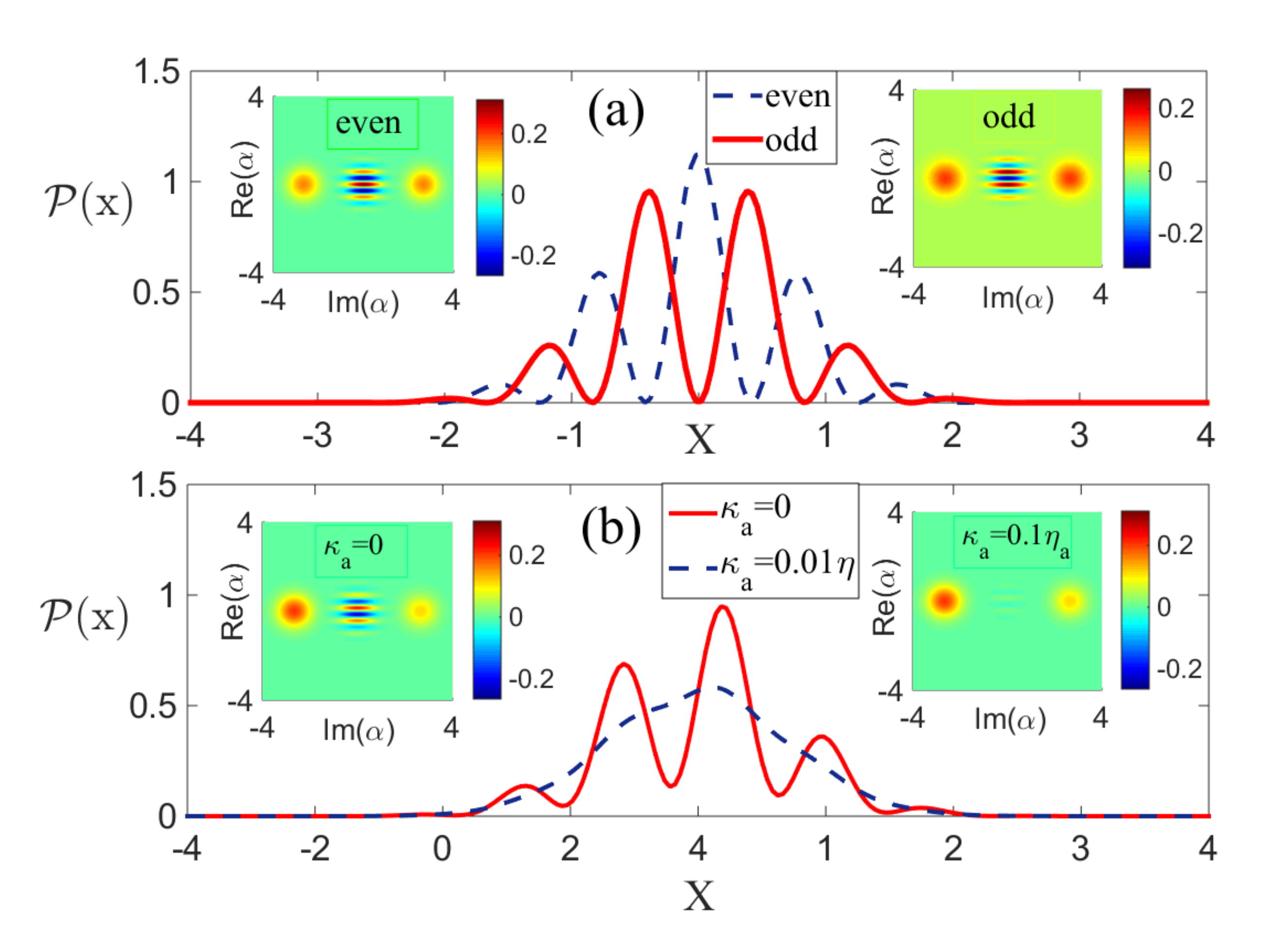}
	\vspace{-1em}
	\caption{(a): Quadrature distributions and the corresponding Wigner functions for the steady states $\rho_{ss}$ without
		single-photon decay. We have chosen the phase in quadrature operator $\phi=0$. The initial states are $|0\rangle$ (even) and $|1\rangle$
		(odd) respectively. (b): The quadrature  distributions and the corresponding Wigner functions for the steady states when the initial states
		are $1/\sqrt{2}(|0\rangle+|1\rangle)$ for the single-photon decay rates being $\kappa_a$=0 and $\kappa_a/\eta_a$=0.1. The asymmetry of the
		quadrature 	distribution corresponds to that of Wigner functions. The other parameters are same to those in Fig. \ref{f:initialstate}.}
	\label{f:quadrature}
\end{figure}

\begin{figure}
	\includegraphics[width=0.5\textwidth]{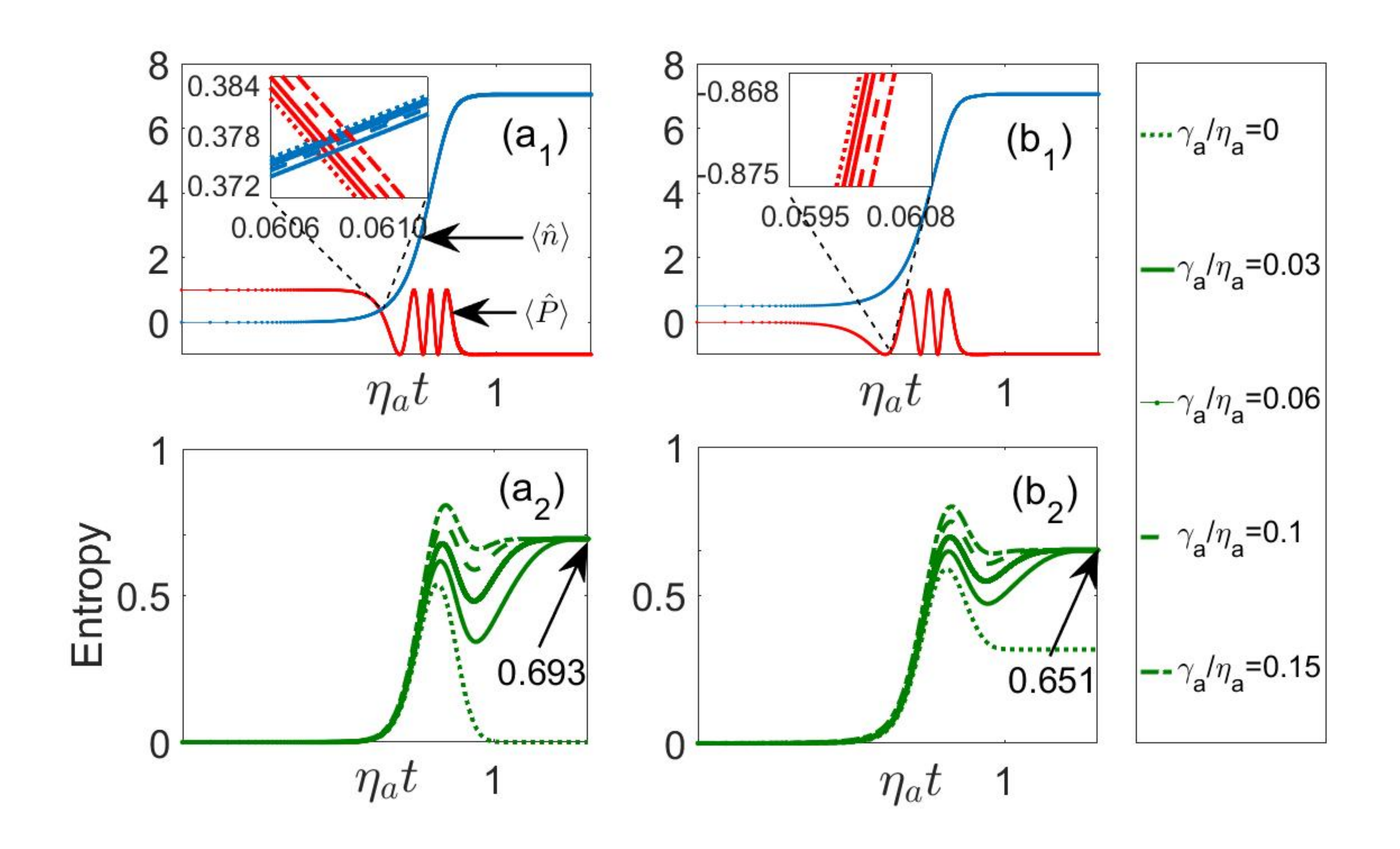}
	\vspace{0em}
	\caption{The evolution of the average photon number $\langle\hat{n}\rangle$ and the parity $\langle\hat{P}\rangle$ are  displayed in ($a_1$) and
		($b_1$) for different $\gamma_a$ and initial states. In ($a_2$) and ($b_2$), the transient dynamics for the entropy are displayed with the decay rates
		 same to those in ($a_1$) and ($b_1$) respectively. In ($a_1$) and ($a_2$), the initial states are the vacuum state $|0\rangle$ while in ($b_1$) and
		 ($b_2$), the initial states are $1/\sqrt{2}(|0\rangle+|1\rangle)$. The other parameters are same to those in Fig. \ref{f:initialstate}. We zoom in on
		 $\langle\hat{n}\rangle$ and $\langle\hat{P}\rangle$ to show the  weeny difference for different decay 	rates with the identical filate in the legend
		 on the right.}
	\label{f:parityn}
\end{figure}
State space which is composed of even Fock states does not link to the one composed of odd Fock states by double-photon driving and decay processes.
Resulting  from the competition between these two double-photon processes, the initial states with deterministic even or odd parity evolve to the one with
the same parity. Here  they are even and odd Schr\"odinger cat states respectively. Whereas if the initial state is a superposition of even and odd Fock states,  the
 steady state would be superimposed of even and odd cat states with different weights.  Another case is the statistical mixture of even and odd Fock states
 in which the ratio of the even and odd cat states in the eigenspace of the steady state would be identical to the initial statistical one. This results from the fact
 that the states in even space and odd space develop separately without single-photon decay or driving process. We display the above arguments by  four
 examples in Fig. \ref{f:initialstate}.
\subsection{Double-photon driven-dissipative process in the presence of single-photon decay}\label{r:oneloss}
One-photon decay leads to mutual exchange of the even and odd  Schr\"odinger cat state since $\hat{a}|\alpha\pm\rangle=\alpha|\alpha\mp\rangle$
which diminishes  the interference fringes in the Wigner functions. The vanishing of the negative regions corresponds to fading away of the interference patterns in
the quadrature distributions. As an example, in Fig. \ref{f:quadrature}, the quadrature distributions are exhibited for the steady states with single-photon decay
rates $\gamma_a=0$ and $0.1\eta_a$ for two different initial states. And the interference structures in the quadrature distributions coincide with the appearance
of negative stripes in the Wigner functions. With increasing of the single-photon decay strength, the negative stripe in the Wigner functions and interference
fringes in the quadrature tends to fade away.  And it can be seen that with the initial state $\frac{1}{\sqrt{2}}(|0\rangle+|1\rangle)$, the steady state become
asymmetry in phase space and quadrature distributions.
\begin{figure}
	\includegraphics[width=0.49\textwidth]{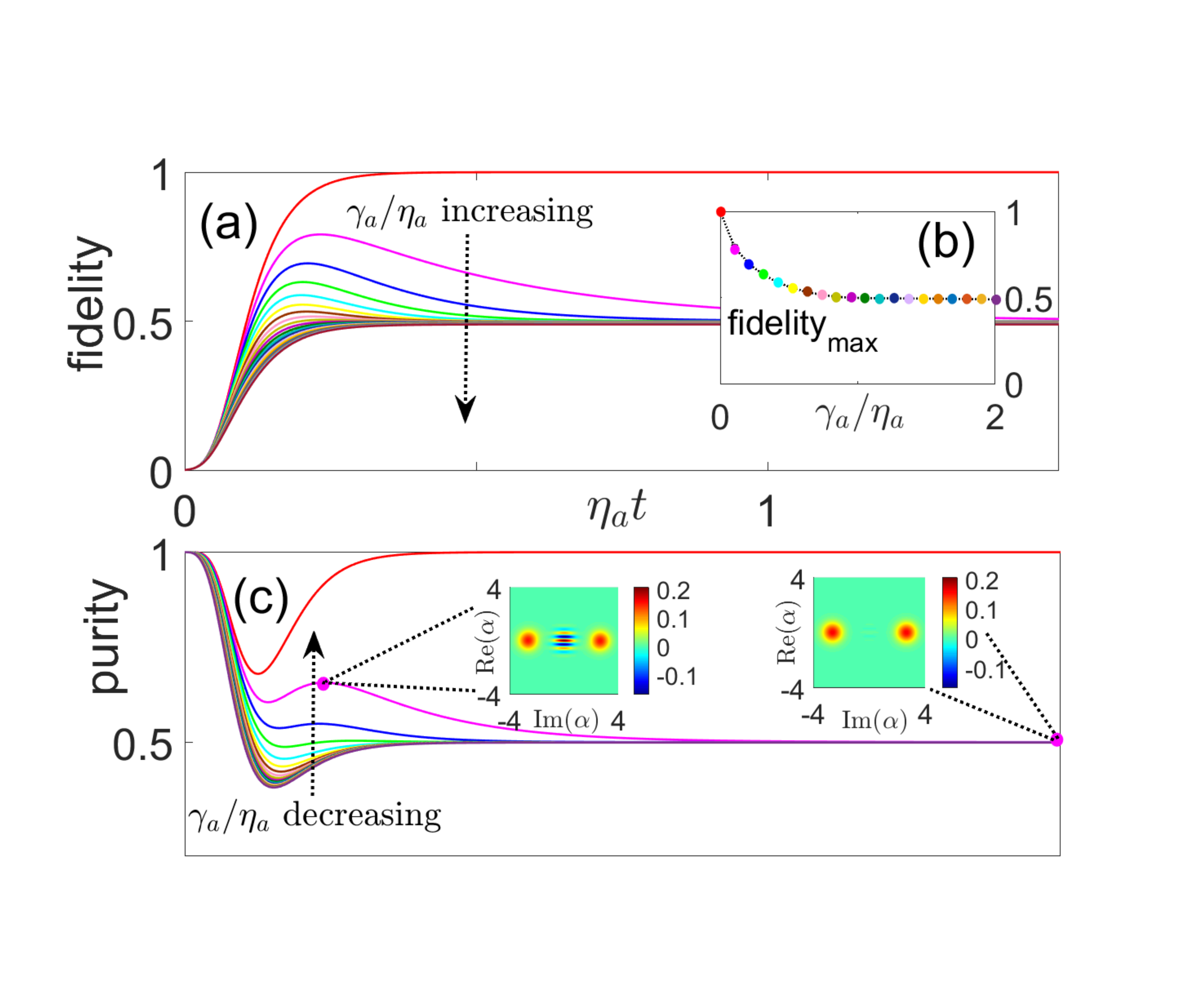}
	\vspace{-1em}
	\caption{ (a): Transient dynamics for the fidelity $|\langle\psi|\psi_{even cat}\rangle|^{2}$ in a range of $\gamma_a/\eta_a$ when $|0\rangle$ is set as
		the initial state.  Panel (b) shows the maximum of the fidelity for the range of $\gamma_a/\eta_a$ during the evolution. (c): The evolution of purity defined as
		$Tr[\rho^2]$ for the same range of $\gamma_a$ in (a). The other parameters are same to those in Fig. \ref{f:initialstate}.}
	\label{f:gamaeffectn}
\end{figure}

Compared with the evolution of average photon number $\langle \hat{n}\rangle$ and parity $\langle \hat{P}\rangle$, the entropy provides more information
to dissect the evolution process and it also reflects some information of the initial state. The entropy can be defined as $S(t)=-Tr[\rho(t)log(\rho(t))]$  where $Tr[\bullet]$
denotes the trace of $\bullet$. We plot
the transient dynamics for $\langle \hat{n}\rangle$, $\langle \hat{P}\rangle$  and the entropy $S(t)$ in Fig. \ref{f:parityn} for several single-photon decay
strengths with two initial states. It can be seen that while the single-photon decay is weak, the entropy would retrace after it drops from the first peak. This behavior
can be interpreted as the competition between the double-photon and single-photon processes, i.e., double-photon process dominates the entropy-decline behavior.
But with time going by, as long as there is nonvanishing single-photon decay, the entropy rises and  converges to an asymptotic value in long time scale. Such an
asymptotic value depends on the initial condition. While the initial state is a vacuum state $|0\rangle$, the asymptotic value would be close to $log(2)$ in the
presence of single-photon decay, the maximum of a two-component statical mixture in the eigenspace of the steady state. Whereas if the initial state is
 $(|0\rangle+|1\rangle)/\sqrt{2}$, the asymptotical value would be less than $log(2)$.  The evolution trajectory for entropy are easier to distinguish than that of
  $\langle\hat{P}\rangle$ and $\langle\hat{n}\rangle$  for different $\gamma$s which may be employed to reflect the strength of single-photon decay.

Even though the Schr\"odinger cat states vanish in the steady state in the presence of single-photon decay, it can still appear during the evolution.  And the life time
of a cat state can be prolonged by increasing  the double-photon dissipation compared to single-photon decay. The appearance  of the Schr\"odinger cat state in the
 presence of single-photon decay is shown in  Fig.\ref{f:gamaeffectn}. As shown in this figure, when $\gamma_a\lesssim\eta_a$, the fidelity can reach more than
0.5 during the evolution. The snap plots of the Wigner functions in Fig. \ref{f:gamaeffectn} indicate the appearance of the cat states. Even though the cat states can
be generated during the evolution, single-photon decay leads  the fidelity approaching 0.5 in long time scale. Here the states may evolve to the statistical mixture of
even and odd Schr\"odinger cat states with identical weights in the eigenspace. This can be confirmed by the purity defined as $Tr[\rho^2]$. When there is single-photon
decay ($\gamma_a>0$), the purity converges  to 0.5 in long time scale. As mentioned above, the entropy converges to $log(2)$ implying  that there are mainly two independent  components(or cats) with the same weights in the steady state. These signatures suggest that the steady state is a statistical mixture of odd and even Schr\"odinger cat states.

\begin{figure}
	\includegraphics[width=0.48\textwidth]{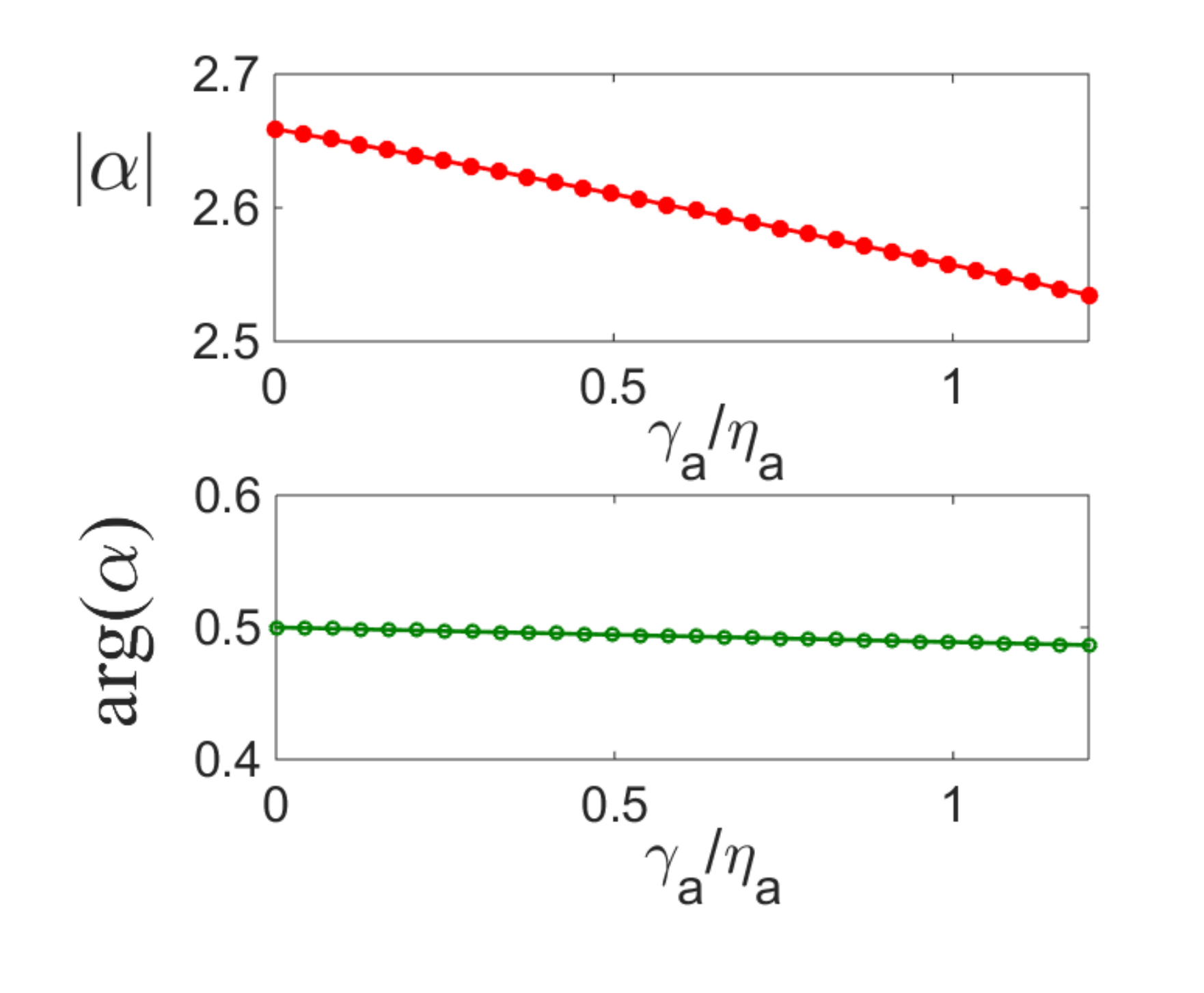}
	\vspace{-1.5em}
	\caption{The influence of the single-photon decay $\gamma_a$ to `$\alpha$' in the steady state. Here `$\alpha$'	denotes the `size' of the Schr\"odinger cat state.
		 The other	parameters are same to those in Fig. \ref{f:initialstate}. }
	\label{f:gamma}
\end{figure}

Then we check the influence of single-photon decay strength to the steady state in detail. The steady state can be determined by solving the equation $\partial_{t}\rho=0$.
 Numerical results in Fig. \ref{f:gamma} indicate that the modulus of  $\alpha$ in the Schr\"odinger cat states decreases linearly as $\gamma_a$ increases. But the
 influence  to the modulus of $\alpha$ is more obvious than that to the angle of $\alpha$. This means larger $\gamma_a$ leads to more degree of reduction of the
 average photon  number in the  steady state since $\langle\pm\alpha|\hat{n}|\pm\alpha\rangle=|\alpha|^2$.

\section{two coupled-modes system}\label{r:twomode}
Based on the one-mode system  discussed above, we now consider the two coupled-modes. Two cases with different kinds of couplings are considered in
following, namely the case with linear coupling  and the one with nonlinear coupling. These two cases can be expressed as
\begin{figure}
	\includegraphics[width=0.48\textwidth]{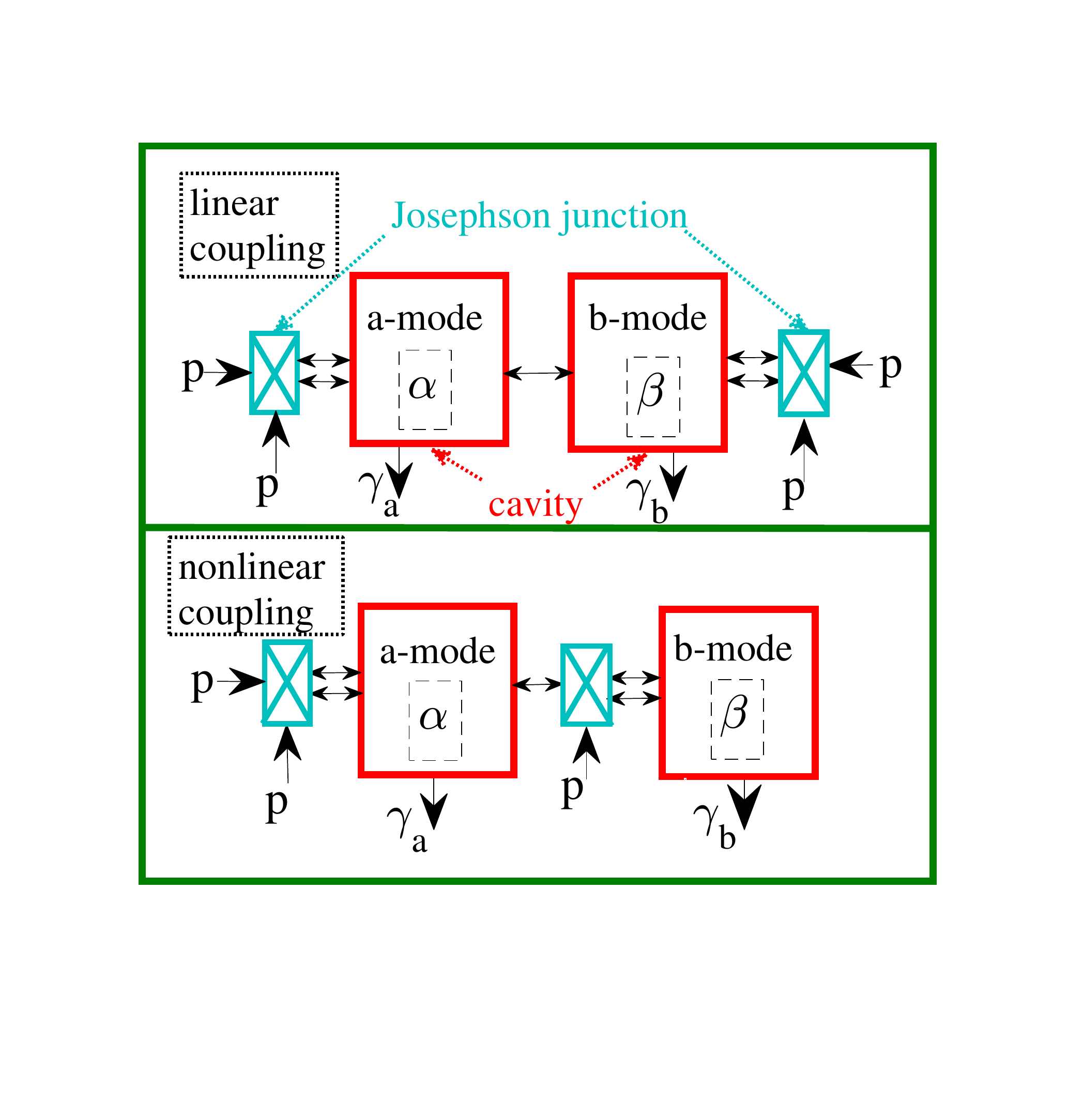}
	\vspace{-1.5cm}
	\caption{The sketches for the cases with linear and nonlinear couplings. The Josephson junctions provide the realization for two-photon processes by four-wave
		mixing process. $\alpha$ and $\beta$ in the red boxes indicate the `size' of the coherent state components which constitute the cat states. `$p$'s denote the
		classical driving beams.}
	\label{f:lnonlconfig}
\end{figure}

\begin{eqnarray}
\begin{aligned}
&linear: g_l(\hat{a}\hat{b}^{\dag}+\hat{a}^{\dag}\hat{b}) \\
&nonlinear:  g_{nl}(\hat{a}\hat{b}^{\dag2}+\hat{a}^{\dag}\hat{b}^2),\\
\end{aligned}\label{Ha}
\end{eqnarray}
where $g_{l(nl)}$ denotes the linear (nonlinear) coupling strength between the two modes. In linear coupling case, we consider two identical systems except
 for the phase of double-photon driving terms which influence the distributions of the Schr\"odinger cat states in phase space (as (\ref{sol}) shows). The nonlinear case is
 similar to the quantum description for second harmonic generation which has been considered in \cite{OA28211}. We investigate  the  Schr\"odinger cat states in the
  coupled modes and correlation between the modes characterized by entanglement and mutual information. Both entanglement and  mutual information reflect
  the correlation between the coupled systems merely through different aspects. They play important roles in quantum information theory\cite{mark,PRA65032314}.
   The sketches for the two coupled modes are shown in Fig. \ref{f:lnonlconfig}.
\subsection{Influence of single-photon decay in the case with linear coupling}\label{lc}
\begin{figure}
	\includegraphics[width=0.48\textwidth]{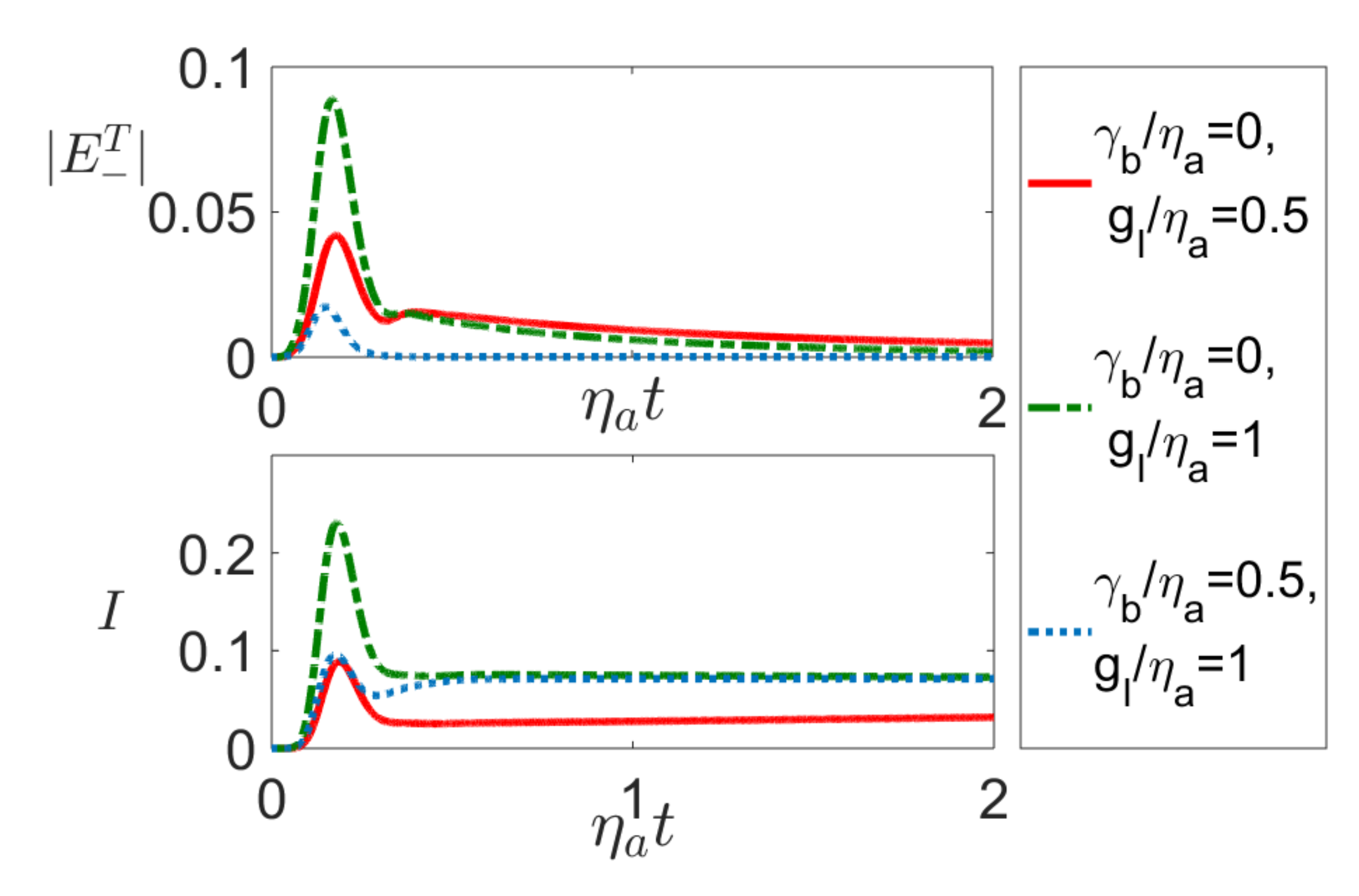}
	\vspace{-1.2em}
	\caption{Transient dynamics for entanglement indicator $|E_-^T|$ and mutual information $I$ in the
		two modes system with linear couplings. We set $\gamma_a/\eta_a$=0.5 in this case and the other parameters are the same as those in
		Fig. \ref{f:initialstate},  except the phase $e^{i3\pi/4}$ in the double-photon driving for $b$-mode.
		The initial states  are $|0\rangle$ for both modes.}
	\label{f:tmlcentg2}
\end{figure}

First, we consider  linear couplings between the two modes as shown in Fig. \ref{f:lnonlconfig}. In this case, the evolution for the total system is described by
\begin{eqnarray}
\begin{aligned}
\partial_t\rho=&-i[\hat{H}_a+\hat{H}_b,\rho]-ig_l[\hat{a}^{\dag}\hat{b}+\hat{a}\hat{b}^{\dag},\rho]\\
&+(\mathcal{L}_{a}^{(1)}+\mathcal{L}_{a}^{(2)}+\mathcal{L}_{b}^{(1)}+\mathcal{L}_{b}^{(2)})\rho.
\end{aligned}
\end{eqnarray}
Here $\rho$ is the density matrix for the total system. $\hat{H}_a$ and $\hat{H}_b$ are Hamiltonians for the two coupled modes. In this case, not only
coherence behavior but also correlation between modes are worth  examining.  For the one mode system, the nonclassical character can be described  by
negative value of the Wigner functions. For the  two coupled-modes, however, the entanglement and mutual information may be interesting to be examined. Since
the appearance of negative eigenvalues for the partial transposed density matrix of a compound system is the necessary condition for nonseparability
\cite{PRL771413,PRA65032314}, the absolute value for the summation of the negative eigenvalues of the partial transposed density matrix can be used as the
entanglement indicator, i.e., the $ |E_-^T|=\sum_{\lambda_i<0}|\lambda_{i}|$, where $\lambda_i$ denotes the negative eigenvalues of the partial
 transposed density matrix. Besides, quantum mutual information, or von Neumann mutual information  provide us with a way  to characterize the information
of  a variable `$A$' in  one system by exploring  its partner `$B$' in another system \cite{Shannon,Cover,RMP80517,PRA91012301,PRL90050401}.
It can be regarded as the quantum version of Shannon  mutual information defined  as $I=S_A+S_B-S_{AB}$ where $S_{A(B)}$ is the entropy for $A(B)$
mode and $S_{AB}$ is that for the compound bipartite system. For comparison, we calculate and plot the dynamics  for the mutual information $I$ along with the
entanglement indicator $|E_-^T|$ in Fig.\ref{f:tmlcentg2} for three examples. In the presence of single-photon decay in one of the coupled modes, the larger
coupling strength leads to sharp changes in  $|E_-^T|$. This can be interpreted as fast exchange of correlations between the two modes caused by larger
 couplings. This corresponds to single-photon decay caused  destruction of nonclassical characters and the  sharp  destruction of the entanglement. The
 stronger the coupling is, the sharper  the change  of the mutual information. Similarly, strong couplings lead to depression of the entanglement at  long time
 scale. Sum up, single-photon decay has negative effect on the entanglement since  $|E_-^T|$ decreases sharply when the single-photon  decay is nonvanishing.
  But the  mutual information  behaves robustly against this decay in the case of linear coupling.

\begin{figure}
\includegraphics[width=0.48\textwidth]{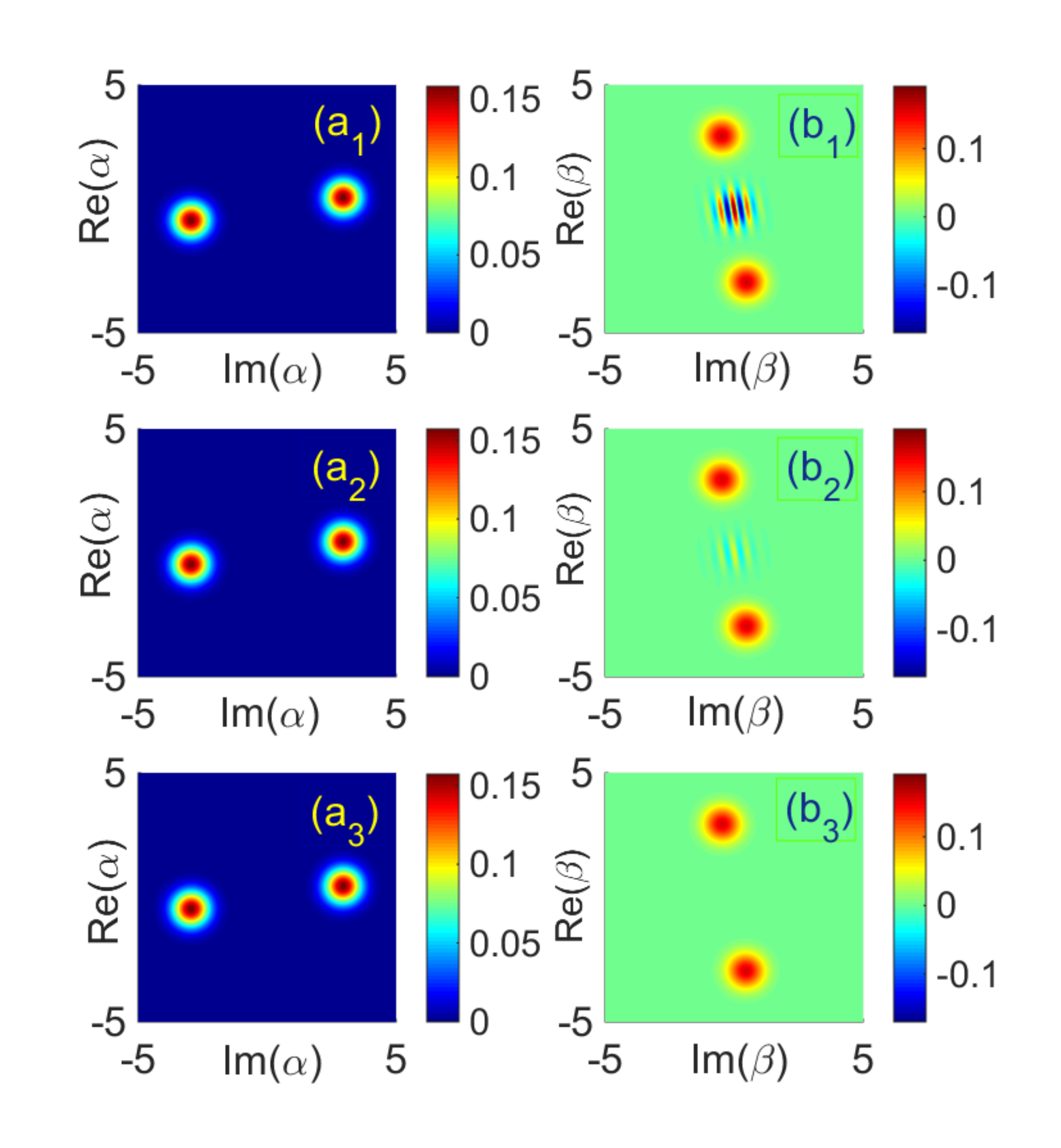}
\vspace{-1em}
\caption{($a_1$), ($a_2$) and ($a_3$): The Wigner functions for $a$-mode in the case with linear couplings at time $\eta_at=2$. ($b_1$), ($b_2$) and ($b_3$):
	The Wigner functions for $b$-mode corresponding to those in ($a_1$), ($a_2$) and ($a_3$) respectively. The other parameters in both modes are the
	same as  those in Fig. \ref{f:tmlcentg2}. }
	\label{f:tmlcwig}
\end{figure}

To explore the effect  of single-photon decay on the coherence, we display the Wigner functions for the coupled modes at scaled time $\eta_at=2$ in
Fig. \ref{f:tmlcwig}. It can be seen that while there is single-photon decay in one of the coupled modes, the strong  linear couplings  lead  to  severe
destruction to the negative regions in the Wigner functions. When the single-photon decay exists in both modes, the negative values in Wigner functions
disappear at long time scales.
\subsection{Influence of single-photon decay in the case with nonlinear coupling}\label{nlc}
\begin{figure}
	\includegraphics[width=0.50\textwidth]{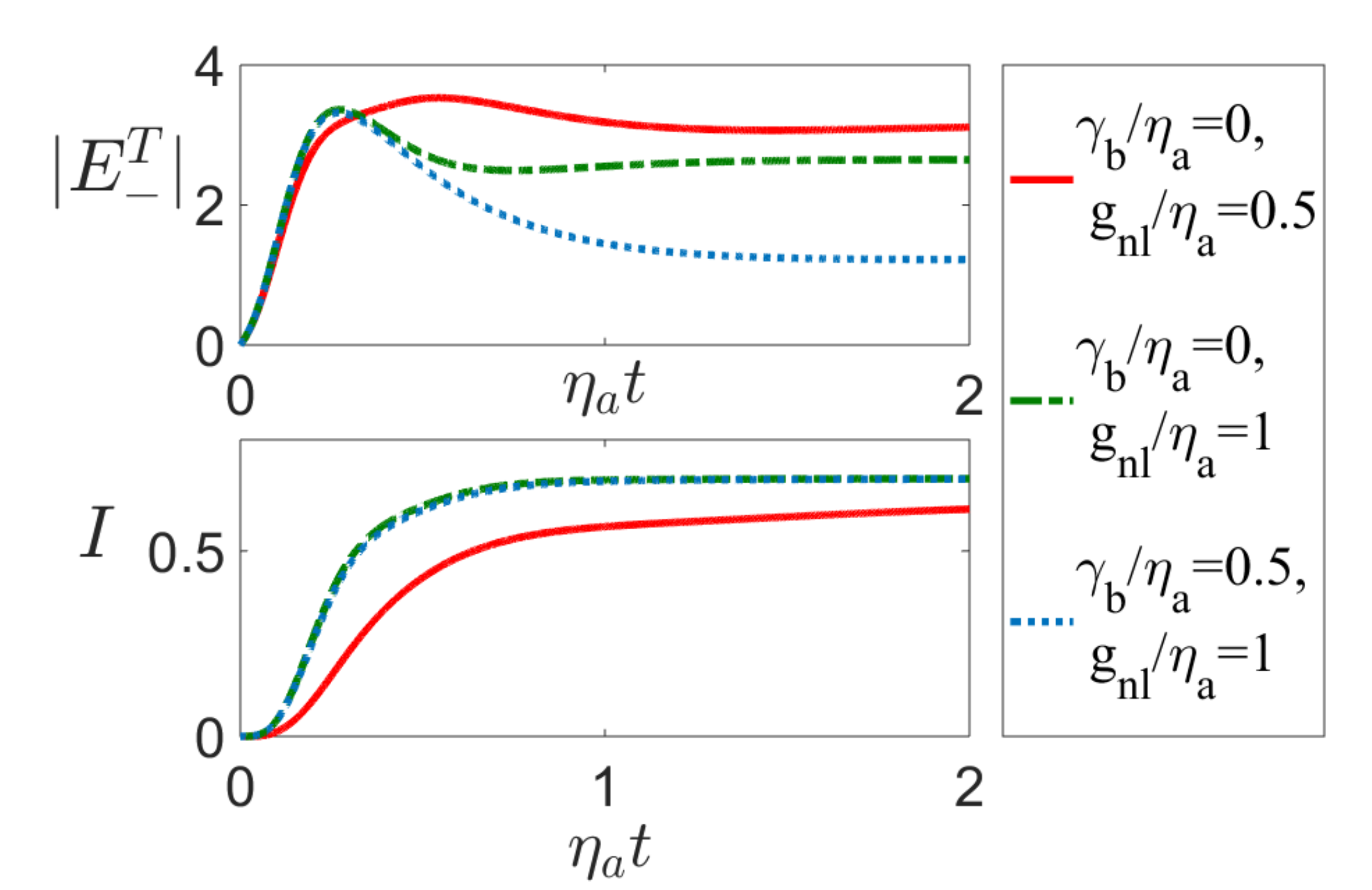}
	\vspace{-1.2em}
	\caption{Transient dynamics for  $|E_-^T|$ and mutual information $I$ in the case with nonlinear couplings. We set $\gamma_a/\eta_a$=0.5
		in this case. The other parameters are the same as those in Fig. \ref{f:initialstate}. The initial state is $|0\rangle$ for both modes.}
	\label{f:tmNlcentg}
\end{figure}
\begin{figure}
	\includegraphics[width=0.52\textwidth]{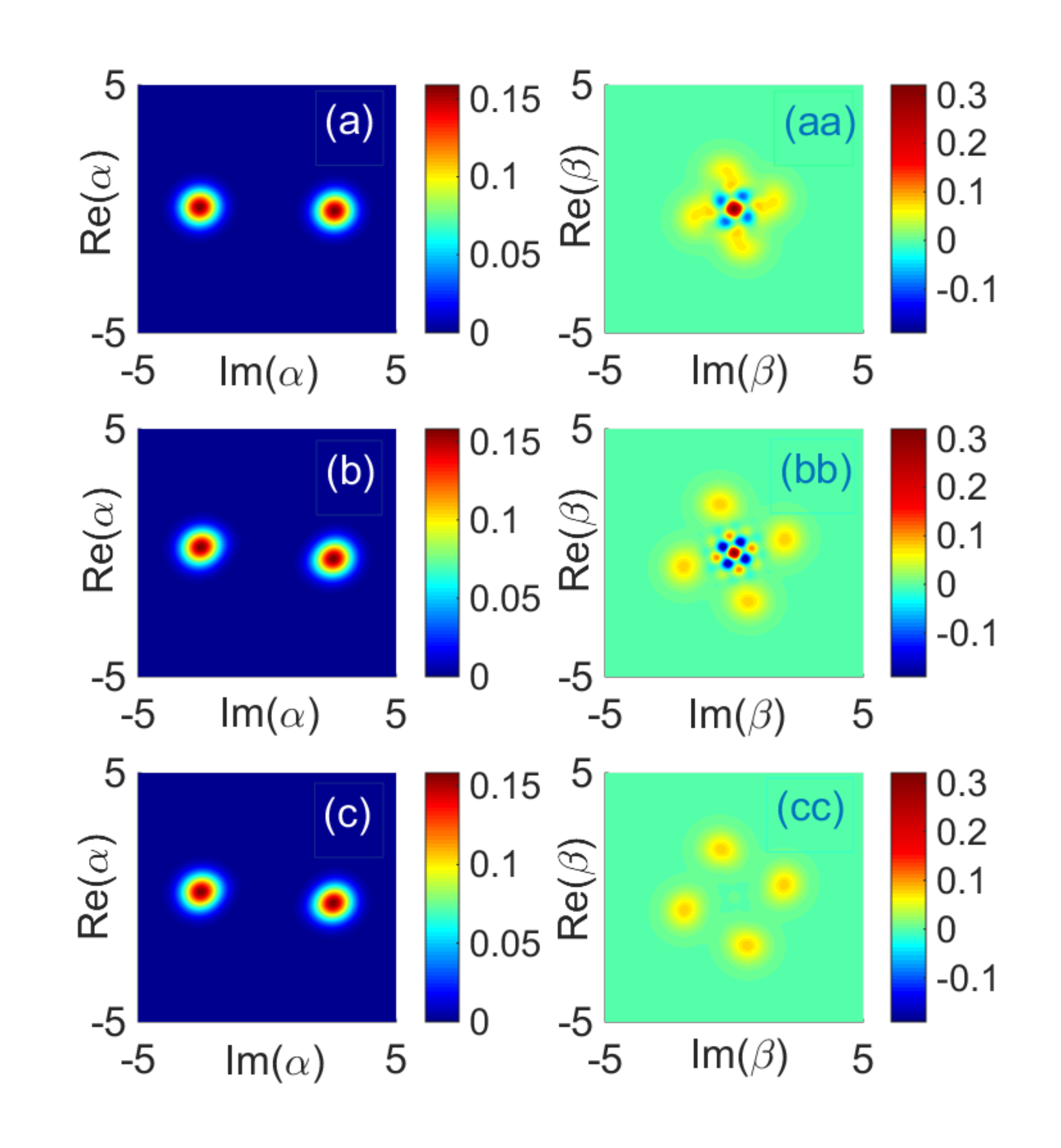}
	\vspace{-1.2em}
	\caption{(a), (b) and (c): Wigner functions for $a$-mode at time $\eta_at=2$ in the two modes system with nonlinear couplings.(aa), (bb)
		 and (cc): Wigner functions for $b$-mode corresponding to (a), (b) and (c), respectively. The other parameters are the same as those
		  in Fig. \ref{f:tmNlcentg}.}
	\label{f:tmNlcwig}
\end{figure}
In the case with nonlinear couplings, we consider a toy model in which the $a$-mode  acts as the double-photon
driving for another mode ($b$-mode, i.e. the sub-converted mode) with double-photon decay. The sketch of such a model is shown in Fig. \ref{f:lnonlconfig}.
The dynamics is governed by the equation,
\begin{figure}
	\includegraphics[width=0.5\textwidth]{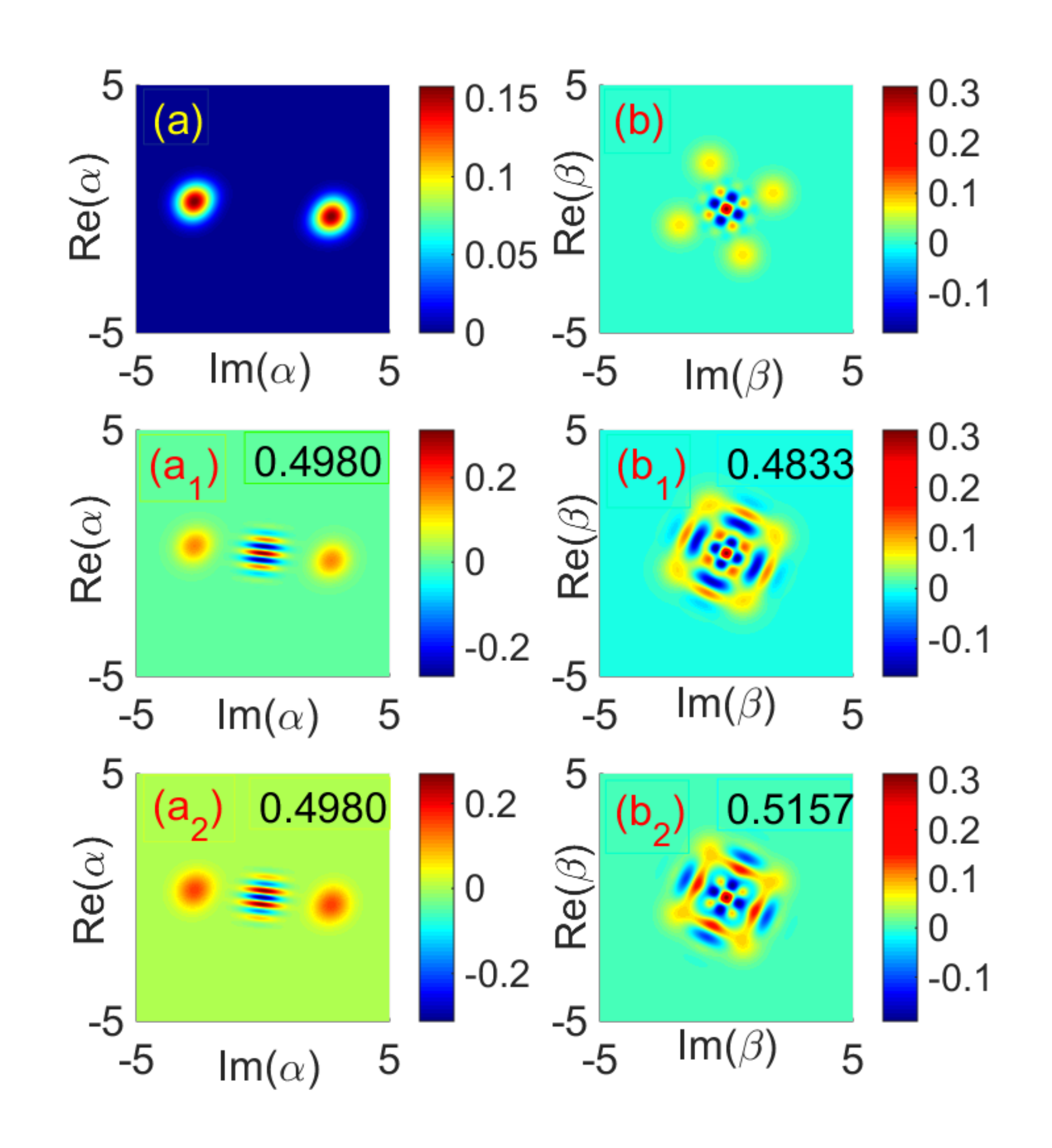}
	\vspace{-1.2em}
	\caption{$(a)$, $(a_1)$ and $(a_2)$ : The Wigner functions for $a$-mode and its two major components in its eigenspace at
		time $\eta_at=3$ with parameters as the  same as in  Fig. \ref{f:tmNlcwig}(b). $ (b)$, $(b_1)$ and $(b_2)$ are those for $b$-mode
		with the same parameters as that in  Fig. \ref{f:tmNlcwig}(bb). The decimals on the panels are the ratios for the components in the
		eigenspaces. The summation of major ratios are  more than 0.99.}
	\label{f:eigenspace}
\end{figure}

\begin{eqnarray}
\begin{aligned}
\partial_t\rho=&-i[\hat{H}_a+\tilde{H}_b,\rho]-ig_{nl}[\hat{a}^{\dag}\hat{b}^2+\hat{a}\hat{b}^{\dag 2},\rho]\\
&+(\mathcal{L}_{a}^{(1)}+\mathcal{L}_{a}^{(2)}+\mathcal{L}_{b}^{(1)}+\mathcal{L}_{b}^{(2)})\rho,\\
\tilde{H}_b=&-\Delta_b\hat{b}^\dagger \hat{b}+\frac{U_b}{2}\hat{b}^{\dag 2}\hat{b}^2.
\end{aligned}
\end{eqnarray}

\begin{figure}
	\includegraphics[width=0.5\textwidth]{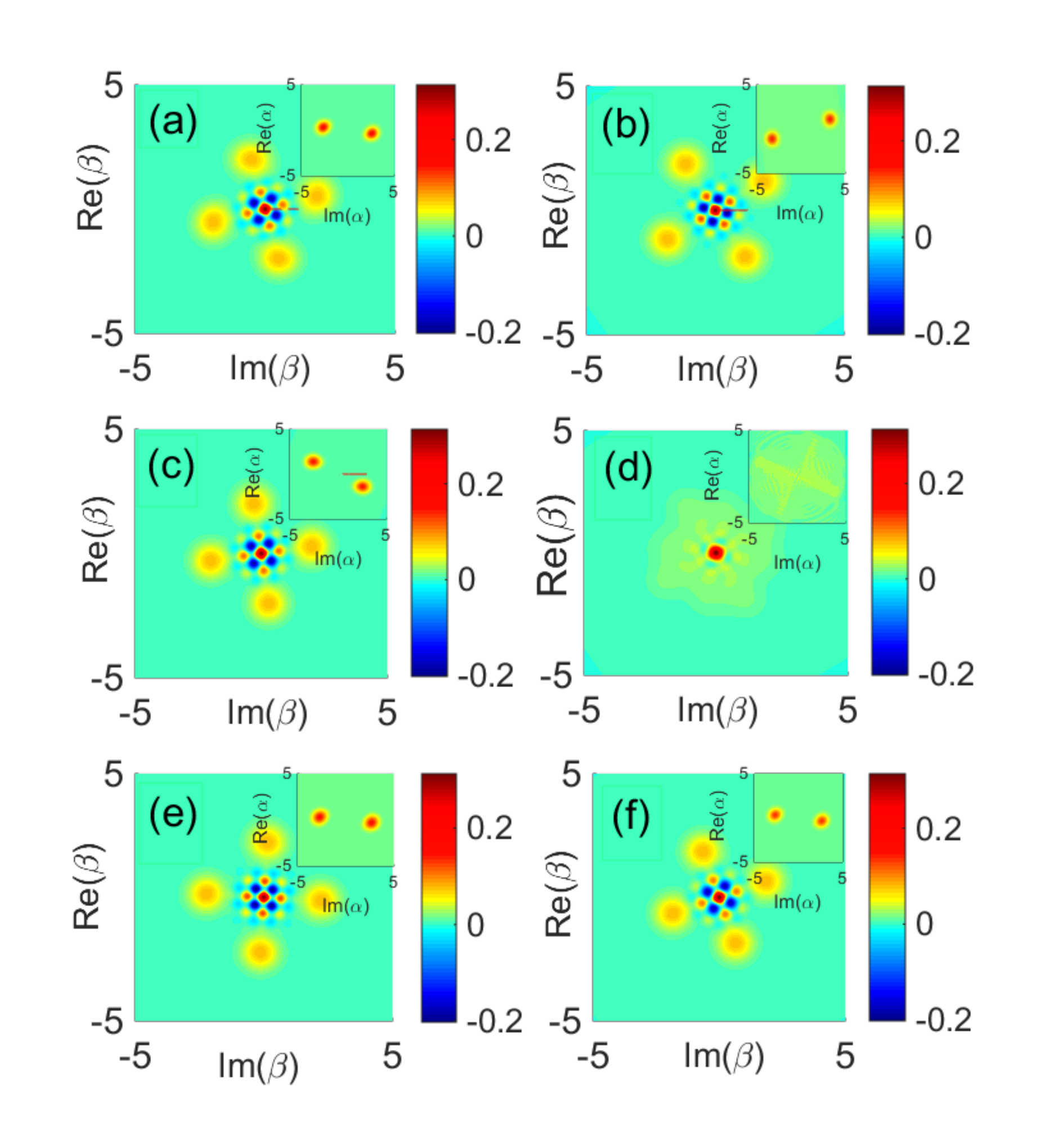}
	\vspace{-1.2em}
	\caption{In (a)-(f) we set: $\gamma_a$=(0.5,0.5,0.5,0.5,0.5,0), $\eta_a$=(1,0,1,0,1,1), $U_a$=(1,1,0,0,1,1), $U_b$=(1,1,1,1,0,1).
		The other parameters are   $G_a$ = $10e^{-i\pi/4}$, $U_b$=1, $\Delta_b$=0, $\eta_b$=1,  $\gamma_b$=0, $g_{nl}$=1.
		Here we choose $|G_a|/10$ as the energy units since both $\eta_a=0$ and $U_a=0$ situations are examined in these examples.
		The  insets   are the Wigner functions for the corresponding $a$-modes with the same color map to $b$-modes.}
	\label{f:blast}
\end{figure}
This can be seen as the cascaded double-photon down conversion scheme. In this case, the entanglement $|E_-^T|$, mutual information $I$, and nonclassical signatures
behave differently with respect  to those in the linear coupling case. We study  the influence of single-photon decay in both modes to $|E_-^T|$, $I$, and coherence by
three examples, and the results are shown  in Fig. \ref{f:tmNlcentg} and Fig. \ref{f:tmNlcwig}. When single-photon decay exists in $a$-mode, with the increasing
of nonlinear coupling strength $g_{nl}$, $|E_-^T|$ would decrease  but $I$ is enhanced. When single-photon decay is added in $b$-mode ($\gamma_b\neq0$), the $|E_-^T|$
decreases  obviously but the mutual information is robust against this decay.

In Fig. \ref{f:tmNlcwig}, intriguing nonclassical signature emerges in the $b$-mode when there is only double-photon decay in this mode. To reveal the intriguing
patterns in both $a$ and $b$-mode, we calculate  their components and show the results  in Fig. \ref{f:eigenspace}. It can be seen that the state of $a$-mode
 is a superposition of even and odd Schr\"odinger cat states. Whereas $b$-mode is superimposed by two four-component cat-like states. The scars  in the
 Wigner functions indicate that  qubits could be encoded  by these states. E.g., the qubits here can be encoded as
$|\alpha\rangle_i \rightarrow|0\rangle_i$ and $|-\alpha\rangle_i \rightarrow|1\rangle_i,$ where $i$ denotes the mode index  and $|\pm\alpha_i\rangle$
are the coherent states. The Schr\"odinger cat states might find potential applications in quantum metrology\cite{Science342607,Nature412712,Science342568,Science3521087,
PRA73023803,PRA78034101,PRA70053813,PRA66023819,PRA66023819}  and  may be valuable in efficient storage and communication for quantum information
 processing \cite{nature448784,pra64052308,science31283,arxiv0509137,Science342607}. This is due to the sensitivity of the states  to the structure change in  phase
 space, for instance, rotation and  shift.

The intriguing patterns in $b$-mode  may be influenced by $a$-mode though coupling. In the following,  we will check the Wigner functions of  $a$-mode and $b$-mode at
time $\eta_at=2$ by varying a set of parameters in $a$-mode in Fig. \ref{f:blast}. We find  that the double-photon driving or photon self-interaction in $a$-mode is necessary
 for the emergence of intriguing checkerboard patterns in $b$-mode. In turn, the pattern in $b$-mode depends on  the parameters of  $a$-mode. This may provide us a method
to estimate system parameters by examining the coupled  two modes.

\begin{figure}
	\includegraphics[width=0.52\textwidth]{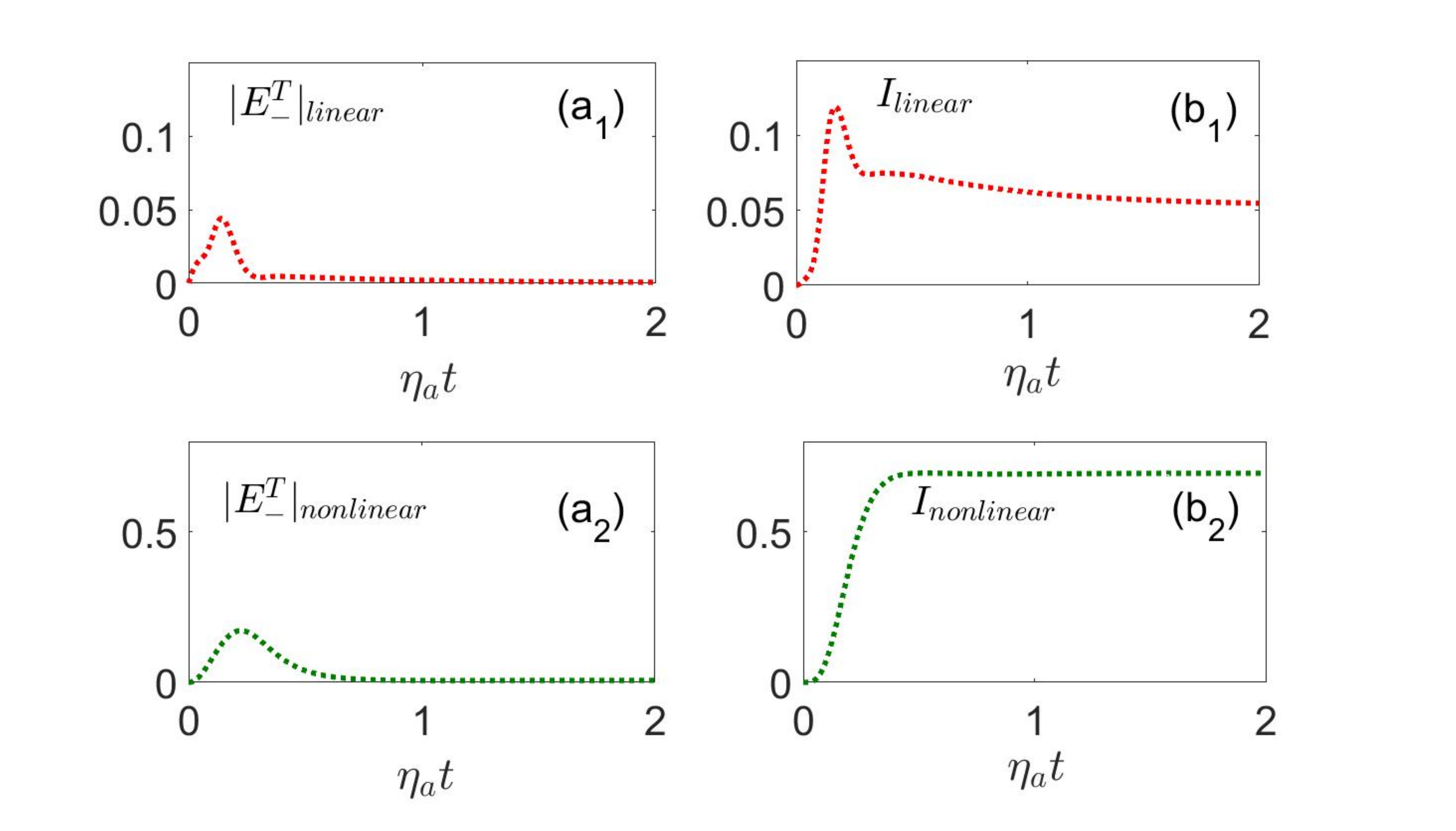}
	\vspace{-1.2em}
	\caption{($a_1$) and ($b_1$) show the entanglement $|E_-^T|_{linear}$ and mutual information $I_{linear}$ respectively in the  case with  linear couplings when
	$\gamma_b/\eta_a=0$ and $g_l/\eta_a=0.5$.  ($a_2$) and ($b_2$) are those in the case with  nonlinear couplings when $\gamma_b/\eta_a=0$ and $g_{nl}/\eta_a=0.5$.
	 The initial state for $b$-mode are taken to be   $\frac{1}{\sqrt{2}}(|0\rangle+|1\rangle)$, while it is the vacuum  state for $a$-mode. The other parameters are the same
	 as those in Fig. \ref{f:tmlcwig} ($a_2$), ($b_2$) for the case with linear couplings and Fig. \ref{f:tmNlcwig}  ($b$), ($bb$) for the case with nonlinear couplings respectively.}
	\label{f:tmentmut}
\end{figure}

\begin{figure}
	\includegraphics[width=0.5\textwidth]{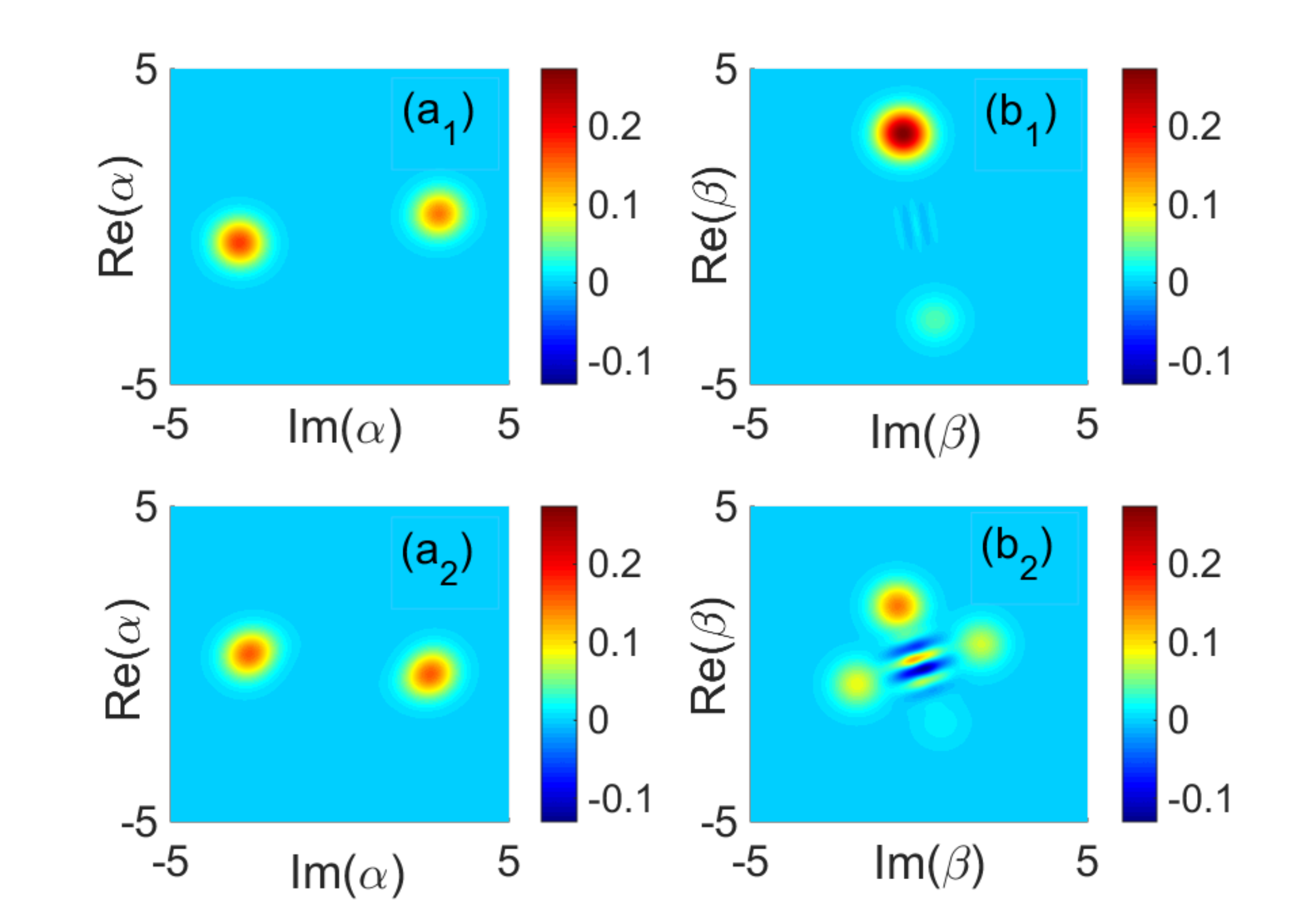}
	\vspace{0em}
	\caption{($a_1$) and ($b_1$) show the Wigner functions for $a$-mode and $b$-mode in the linear coupling case at
		time $\eta_at=2$. ($a_2$) and ($b_2$) are the Wigner functions in the nonlinear coupling case. The initial states and parameters in ($a_1$), ($a_2$), ($b_1$)
		and ($b_2$) are the  same as those in Fig. \ref{f:tmentmut}. In the linear coupling case, the two peaks in the Wigner function in
		$a$-mode are asymmetric which are more obvious than that in the nonlinear coupling case. }
	\label{f:tminitial}
\end{figure}
\subsection{Schr\"odinger cat states in the two-mode system with two different initial states}\label{twoinitial}
We calculate and show  the dynamics of entanglement $|E_-^T|$ and mutual information $I$ in the two modes system for two different initial states in Fig. \ref{f:tmentmut}.
Compared to those in Fig. \ref{f:tmlcentg2} and Fig. \ref{f:tmNlcentg}, we find that with both  linear and nonlinear couplings,  $|E_-^T|$ and $I$ have been suppressed.  However,
 with nonlinear coupling, the entanglement has  been suppressed  more obviously compared to those in  Fig. \ref{f:tmNlcentg}, while the mutual information is robust in this case.
\begin{figure}
	\includegraphics[width=0.52\textwidth]{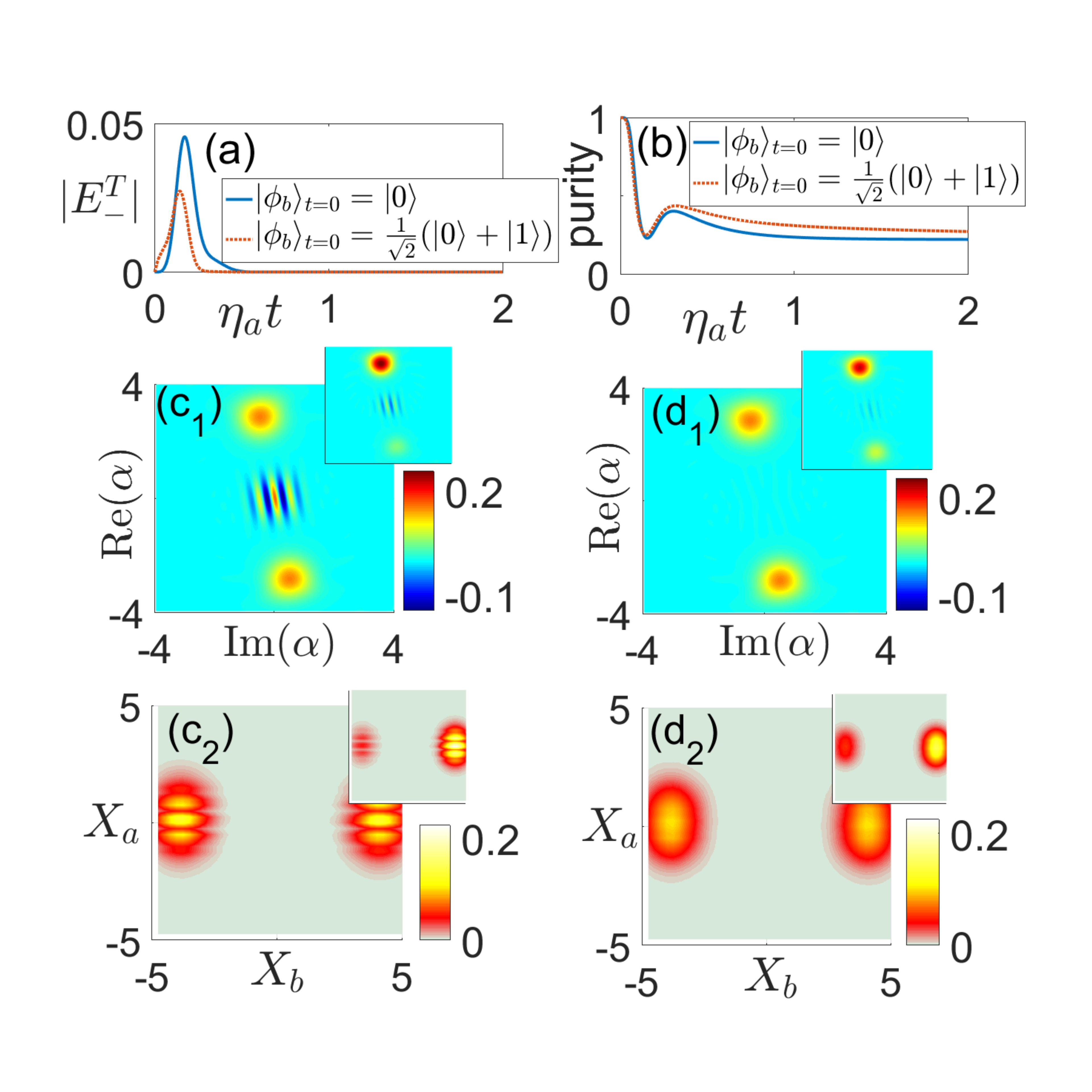}
	\vspace{0em}
	\caption{(a) and (b) are the evolution of the entanglement indicator $|E_-^T|$ and the purity with linear couplings for different initial states (solid curve:
		$|\phi_b\rangle=|0\rangle$; dashed curve:  $|\phi_b\rangle=(|0\rangle+|1\rangle)/\sqrt{2}$. ($c_1$) and ($c_2$) are
		the Wigner functions for $b$-mode and the corresponding joint quadrature distributions at time $\eta_at$=0.4 when $|\phi_b\rangle=|0\rangle$.
		 ($d_1$) and ($d_2$) are the Wigner functions for $b$-mode and the  corresponding joint quadrature distributions at time $\eta_at$=2. The insets
		  are those with initial states of $b$-mode:  $|\phi_b\rangle=(|0\rangle+|1\rangle)/\sqrt{2}$ as shown in (a) and (b) while $|\phi_a\rangle=|0\rangle$.
		  We have set  $\gamma_a/\eta_a$=$\gamma_b/\eta_a$=0.1 and $g_l/\eta_a$=1 and the other parameters are the same as those in Fig.~\ref{f:tmlcentg2}.}
	\label{f:tmlqurd}
\end{figure}

Next we explore  the change of coherence for two different initial states. Compared to Fig.\ref{f:tmlcwig} and Fig.\ref{f:tmNlcwig},  we examine the Wigner
functions at time $\eta_at=2$ in Fig. \ref{f:tminitial} with the same parameters as in Fig. \ref{f:tmentmut}. The negative region in the Wigner functions  approximately
disappears in $a$-mode and small area with negative value of Wigner function left in $b$-mode in the linear coupling case. The influence on $b$-mode distribution is
similar to that in the one-mode case as shown in Fig.\ref{f:quadrature}: the distribution becomes asymmetric in phase space. The distribution for $a$-mode has
also become asymmetric. With nonlinear couplings, compared to those in	Fig.\ref{f:tmNlcwig} ($b$) and ($bb$), the interference fringes in Fig.\ref{f:tminitial}
 ($a_2$) and ($b_2$) also changes obviously in $b$-mode more than that in $a$-mode. This may result from that the nonlinear coupling terms act like single-photon decay for
$a$-mode. Compared to the symmetric pattern in the Wigner function (see Fig. \ref{f:tmNlcwig}), the asymmetric signatures in Fig. \ref{f:tminitial} ($b_2$) too,
 resemble that in the one-mode case.
\begin{figure}
	\includegraphics[width=0.51\textwidth]{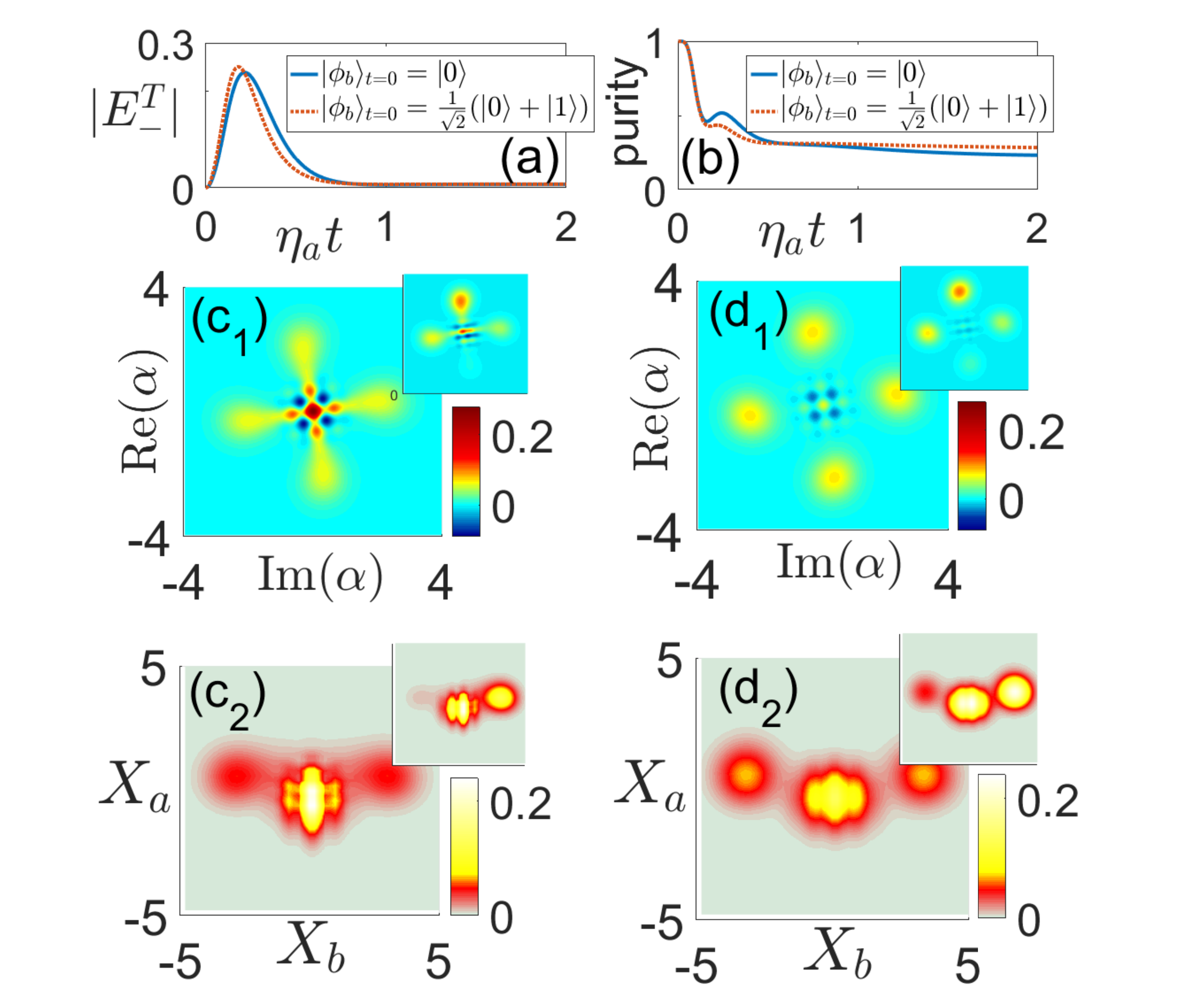}
	\vspace{0em}
	\caption{(a) and (b) are the evolution of  $|E_-^T|$ and the purity in the case with  nonlinear coupling for different initial states (solid curve:
	$|\phi_b\rangle=|0\rangle$; dashed curve:  $|\phi_b\rangle=(|0\rangle+|1\rangle)/\sqrt{2}$. ($c_1$) and ($c_2$) are the Wigner functions
 for $b$-mode and the corresponding joint quadrature distributions at time $\eta_at$=0.4 when $|\phi_b\rangle=|0\rangle$.  ($d_1$) and ($d_2$)
are the Wigner functions for $b$-mode and the corresponding joint quadrature distributions at time $\eta_at$=2. The insets are those with initial
states of $b$-mode: $|\phi_b\rangle=(|0\rangle+|1\rangle)/\sqrt{2}$  as shown in (a) and (b).	We have set $\gamma_a/\eta_a$=$\gamma_b/\eta_a$=0.1
and $g_{nl}/\eta_a$=0.5 and the other parameters are the  same as  those in Fig.~\ref{f:tmNlcentg}.}
	\label{f:tmnlqurd}
\end{figure}
\subsection{The joint quadrature distribution in the two coupled-modes cases}\label{twoquard}
In the two coupled-modes case,  the joint quadrature distribution defined by $\langle X_a,X_b|\rho|X_a,X_b\rangle$ provides with the other aspects to witness
the nonclassical characters of the cat states. Here $X_a$ and $X_b$ are the quadrature operators of $a$-mode and $b$-mode in (\ref{qp}) with $\phi=0$
and $\rho$  being the total density matrix of the coupled system. We show the Wigner functions and the corresponding joint quadrature distributions at two
different times in the linear and nonlinear coupling cases in Fig.\ref{f:tmlqurd} and Fig.\ref{f:tmnlqurd} respectively.  While there is negative regions
in the Wigner functions,  interference fringes also appear in the joint quadrature distributions.  When the Wigner function is  symmetric (asymmetric),
the joint quadrature   distribution is also  symmetric (asymmetric). Considering the interference fringes in the quadrature distributions in one-mode case,
 we can see that measurement of the quadrature $X_a$ may project the system to a Schr\"odinger cat state. The joint quadrature distribution has been used to
 explore Einstein-Podolsky-Rosen  paradox, see  Refs.~\cite{pr47777,prl114100403,prl602731,rmp811727,josab32a82}. The purity trends to 0.25 which
 hints that  $a$- and $b$-mode approaches to a statistical  mixture of even and odd cat states at  long time scale.
\section{summary}\label{r:sum}
In summary, we have explored  the Schr\"odinger cat states in two-photon driven dissipative systems of  one mode and two cases of coupled-modes. In one mode
case, the steady state is a superposition of  coherent states, the weight of each  component in  the state is  determined by the system parameters. Compared  to average
photon number and parity, entropy can characterize the dynamics more precisely. Single-photon decay leads to vanishing of the interference fringes in the quadrature
 distributions which corresponds to the disappearance of negative regions in the Wigner functions. But the Schr\"odinger cat states can still  emerge  in the presence of
 weak single-photon decay.  The  single-photon decay slightly decreases  the modulus of  these  coherent states in  the Schr\"odinger cats when the system
reaches the steady state. Whereas the double-photon dissipation can prolong the lifetime of the Schr\"odinger cats.

In the two-mode systems with both  linear and nonlinear couplings, single-photon decay suppresses the entanglement at long time scales.  And this deleterious effect
 of the single-photon decay on the entanglement is severer in the linear coupling case than that in the nonlinear coupling case. However the coupling  benefits the mutual
  information, and intriguing nonclassical  features appear in the low frequency mode in the nonlinear coupling case. But they are quite fragile  against the single-photon
  decay. Such cat states not only provide us with candidates to explore the boundary  between quantum and classical world,  but also  can be applied in precision
   measurement and quantum information processing.
\section*{ACKNOWLEDGMENTS}
This work is supported by the National Natural Science Foundation of China (Grant No. 11534002 and 61475033).

\renewcommand\refname{Reference}
\bibliographystyle{plain}
\bibliography{Thesis}
\end{document}